\documentstyle[prl,aps,preprint]{revtex}

\begin{document}

\bigskip\ 

\begin{center}
{\bf S-DUALITY FOR 2-d GRAVITY}

\bigskip\ 

\smallskip\ 

J. A. Nieto\footnote[1]{%
nieto@uas.uasnet.mx}

\smallskip\ 

{\it Facultad de Ciencias F\'{\i}sico-Matem\'{a}ticas de la Universidad
Aut\'{o}noma}

{\it de Sinaloa, 80010 Culiac\'{a}n Sinaloa, M\'{e}xico}

\bigskip\ 

\bigskip\ 

{\bf Abstract}
\end{center}

We investigate the analogue of S-duality for 2-d gravity. Our analysis is
based in a partition function associated to the Katanaev-Volovich type
action for 2-d gravity. We find a S-dual 2-d gravitational action which is
related to the original 2-d gravitational action by strong-weak duality
transformation.

\bigskip\ 

PACS: 04.60.-m; 04.65.+e; 11.15.-q; 11.30.Ly

July, 2001

\newpage

\noindent {\bf 1.- INTRODUCTION}

\smallskip\ 

The possibility to associate the analogue of S-duality to Einstein
gravitational theory in four dimensions has been pursuing by a number of
authors [1]-[6]. The picture which emerge from these works is that although
technically one can follow a similar procedure as in the case of Abelian
gauge theory and non-Abelian gauge theories [7]-[16] the final gravitational
theory does not possess an exact strong-weak duality symmetry, as in the
case of Yang-Mills theory. The idea can be realized as an exact symmetry,
however, for linearized gravity [6] essentially because such a theory is
indeed an Abelian gauge theory. From this scenario for linearized gravity
arises the interesting possibility to have a strong coupling phase and
small-large duality for the cosmological constant [6]. Linearized gravity
may also be used as an inspiration to study the strong coupling limit for
gravity assuming the possibility of some decompactified dimension. In this
case, it seems that the strong coupling limit for gravity in four dimensions
is a (4,0) conformal field theory in six dimensions [17]-[18].

In this work, we show that 2-d gravity also possesses an exact strong-weak
duality symmetry. Specifically, we derive a S-dual action for 2-d gravity
which is connected to the original 2-d gravitational action by strong-weak
duality transformation. Our work may be of physical interest for at least
two reasons. First, it might be helpful to understand some aspects of
S-duality for gravity in four dimensions and second it might be useful to
understand some aspects of 2-d gravity itself. Due to the fact that 2-d
gravity is deeply related to string theory [19] and black hole physics [20]
the present work may also be of special physical interest in such theories.

The plan of this work is as follows: In section 2, we mention some general
aspects of 2-d gravity and in particular we discuss a 2-d gravity as a gauge
theory. In section 3, we show that it is possible to understand 2-d gravity
as an Abelian gauge theory. In section 4, we apply standard techniques to
derive the S-dual action for 2-d gravity. Finally, in section 5, we make
some comments.

\noindent {\bf 2.- 2-d GRAVITY AS A GAUGE THEORY}

\smallskip\ 

It is a fact that interest in studying 2-d gravity has been growing in the
last few years. Perhaps the main reason for this growing interest is the
deeply connection between quantum 2-d gravity, string theory and black holes
physics in two dimensions. One of the hopes is that 2-d gravity may help to
shed some light about quantum gravity in four dimensions. There are a
variety of models of 2-d gravity (see [21] and references therein). The
simplest consistent theories seem to be the models of Jackiw and Teitelboim
[22] and Callan, Giddings, Harvey and Strominger [20]. In fact, it has been
shown that these two models can be derived through a dimensional reduction
from a Chern-Simons theory [23]. Another interesting proposal is the one due
to Katanaev and Volovich [24], which proposed action is quadratic in the
curvature and in the torsion. Kummer [25] and Hehl [26] have shown that the
action of Katanaev and Volovich has a very interesting black holes solutions.

Here, in order to discuss the analogue of S-duality for 2-d gravity we shall
consider the de Sitter version of Katanaev-Volovich action, which seems to
be due to Solodukhin [27]. This author showed that the equations of motion
derived from such an action are exactly integrated with asymptotic de Sitter
solution which for some assumptions corresponds to charged black holes
solution.

Let us consider a SO(1,2) one form gravitational gauge field $\omega =\omega
_\mu dx^\mu =\frac 12\omega _\mu ^{AB}J_{AB}dx^\mu $ in two dimensions,
where $J_{AB}=-J_{BA}$ are the generators of SO(1,2) and $\mu =0,1$. We
shall split $\omega _\mu ^{AB}$ as a SO(1,1) connection $\omega _\mu ^{ab}$
and the $\omega _\mu ^{2a}=e_\mu ^a$ {\it Zweibein} field, with $a,b=0,1$.
Thus, the de Sitter curvature

\begin{equation}
F_{\mu \nu }^{AB}=\partial _\mu \omega _\nu ^{AB}-\partial _\nu \omega _\mu
^{AB}+\omega _\mu ^{AC}\omega _{\nu C}^{\quad B}-\omega _\nu ^{AC}\omega
_{\mu C}^{\quad B}  \eqnum{1}
\end{equation}
leads to the Macdowell-Mansouri type curvature

\begin{equation}
F_{\mu \nu }^{ab}=R_{\mu \nu }^{ab}+\Sigma _{\mu \nu }^{ab}  \eqnum{2}
\end{equation}
and

\begin{equation}
F_{\mu \nu }^{2a}=\partial _\mu e_\nu ^a-\partial _\nu e_\mu ^a+\omega _\mu
^{ac}e_{\nu c}-\omega _\nu ^{ac}e_{\mu c},  \eqnum{3}
\end{equation}
where

\begin{equation}
R_{\mu \nu }^{ab}=\partial _\mu \omega _\nu ^{ab}-\partial _\nu \omega _\mu
^{ab}+\omega _\mu ^{ac}\omega _{\nu c}^{\quad b}-\omega _\nu ^{ac}\omega
_{\mu c}^{\quad b}  \eqnum{4}
\end{equation}
is the 2-d curvature and

\begin{equation}
\Sigma _{\mu \nu }^{ab}=e_\mu ^ae_\nu ^b-e_\nu ^ae_\mu ^b.  \eqnum{5}
\end{equation}
Of course,

\begin{equation}
T_{\mu \nu }^a\equiv F_{\mu \nu }^{2a}  \eqnum{6}
\end{equation}
can be identified with the torsion.

Let us consider the action

\begin{equation}
S=\frac 1{4\alpha ^2}\int d^2x\sqrt{-g}{}g^{\mu \alpha }g^{\nu \beta
}{}F_{\mu \nu }^{AB}{}F_{\alpha \beta }^{CD}{}\eta _{AC}\eta _{BD}, 
\eqnum{7}
\end{equation}
where $g_{\mu \nu }=$ $e_\mu ^ae_\nu ^b\eta _{ab},$ $g=\det (g_{\mu \nu }),$ 
$(\eta _{AB})=(-1,1,1)$ is the Killing-Cartan metric associated to SO(1,2)
and $\alpha ^2$ is a dimensionless coupling constant.

Using (2)-(6), one may develop the action (7) to obtain

\begin{equation}
\begin{array}{c}
S=\frac{1}{4\alpha ^{2}}\int d^{2}x{}\sqrt{-g}(g^{\mu \alpha }g^{\nu \beta
}{}R_{\mu \nu }^{ab}{}R_{\alpha \beta }^{cd}{}\eta _{ac}\eta _{bd}+g^{\mu
\alpha }g^{\nu \beta }{}T_{\mu \nu }^{a}{}T_{\alpha \beta }^{b}{}\eta _{ab}
\\ 
\\ 
+2g^{\mu \alpha }g^{\nu \beta }{}\Sigma _{\mu \nu }^{ab}{}R_{\alpha \beta
}^{cd}{}\eta _{ac}\eta _{bd}+g^{\mu \alpha }g^{\nu \beta }{}\Sigma _{\mu \nu
}^{ab}{}\Sigma _{\alpha \beta }^{cd}{}\eta _{ac}\eta _{bd}),
\end{array}
\eqnum{8}
\end{equation}
where we consider that $\eta _{22}=1.$ In order to compare the action (8)
with previous actions which are quadratic in torsion and in curvature, let
us introduce the completely antisymmetric quantities $\varepsilon ^{\mu \nu
} $ and $\varepsilon ^{ab}$, with $\varepsilon ^{01}=1$ and $\varepsilon
_{01}=1$, associated to the space-time and the gauge group SO(1,1)
respectively. We have

\begin{equation}
\varepsilon ^{\mu \nu }\varepsilon _{ab}=e(e_{a}^{\mu }e_{b}^{\nu
}-e_{a}^{\nu }e_{b}^{\mu }),  \eqnum{9}
\end{equation}
where $e$ is the determinant of $(e_{\mu }^{a})$. Introducing the density
tensors $\epsilon ^{\mu \nu }=-\frac{1}{e}\varepsilon ^{\mu \nu }$ and $%
\epsilon _{\mu \nu }=e\varepsilon _{\mu \nu }$ and the completely
antisymmetric quantities $\epsilon ^{ab}=-\varepsilon ^{ab}$ and $\epsilon
_{ab}=\varepsilon _{ab}$ the relation (9) gives

\begin{equation}
\epsilon ^{\mu \nu }\epsilon _{ab}=-(e_{a}^{\mu }e_{b}^{\nu }-e_{a}^{\nu
}e_{b}^{\mu })  \eqnum{10}
\end{equation}
and therefore by virtute of (5) we find that

\begin{equation}
\Sigma _{ab}^{\mu \nu }=-\epsilon ^{\mu \nu }\epsilon _{ab}.  \eqnum{11}
\end{equation}
Thus, using (11) we find that (8) becomes

\begin{equation}
\begin{array}{c}
S=\frac{1}{4\alpha ^{2}}\int d^{2}xe(-\frac{1}{2}\epsilon ^{\mu \nu }R_{\mu
\nu }^{ab}{}\epsilon ^{\alpha \beta }R_{\alpha \beta ab}-\frac{1}{2}\epsilon
^{\mu \nu }T_{\mu \nu }^{a}{}\epsilon ^{\alpha \beta }T_{\alpha \beta a} \\ 
\\ 
-2\epsilon ^{\mu \nu }\epsilon _{ab}R_{\mu \nu }^{ab}-2(\epsilon ^{\mu \nu
}\epsilon _{\mu \nu })),
\end{array}
\eqnum{12}
\end{equation}
where we also used the fact that $e^{2}=-g,$ $\epsilon ^{\mu \nu }\epsilon
^{\alpha \beta }=-(g^{\mu \alpha }g^{\nu \beta }-g^{\mu \beta }g^{\nu \alpha
}),$ $\epsilon ^{ab}\epsilon _{ab}=-2$ and $\epsilon _{\mu \nu }=e_{\mu
}^{a}e_{\nu }^{b}\epsilon _{ab}.$

Our final goal is to write (12) in abstract notation. For this purpose we
shall use the differential form notation $R^{ab}=\frac{1}{2}R_{\mu \nu
}^{ab}dx^{\mu }\wedge dx^{\nu }$ and $T^{a}=\frac{1}{2}T_{\mu \nu
}^{a}dx^{\mu }\wedge dx^{\nu }$, with $dx^{\mu }\wedge dx^{\nu }=-dx^{\nu
}\wedge dx^{\mu }$. We shall also use the Hodge dual $^{\ast }R^{ab}=\frac{1%
}{2}\epsilon ^{\mu \nu }R_{\mu \nu }^{ab}$ and $^{\ast }T^{a}=\frac{1}{2}%
\epsilon ^{\mu \nu }T_{\mu \nu }^{a}$. Note that $dx^{\mu }\wedge dx^{\nu
}=-e\epsilon ^{\mu \nu }dx^{0}\wedge dx^{1}.$ Thus, we find that (12) can be
written as

\begin{equation}
S=\frac 1{\alpha ^2}\int_{M^2}(\frac 12TrR\wedge ^{*}R+\frac 12TrT\wedge
^{*}T+R+\frac 12\epsilon _{ab}e^a\wedge e^b).  \eqnum{13}
\end{equation}
Here $M^2$ is a two dimensional manifold, $R=\frac 12\epsilon _{ab}R_{\mu
\nu }^{ab}dx^\mu \wedge dx^\nu $ and $e^a=e_\mu ^adx^\mu .$ We recognize in
the third and fourth terms of (13) the Euler number and the cosmological
constant term respectively. In fact, rescaling the {\it Zweibein} field $%
e_\mu ^a$ in the form $\lambda e_\mu ^a$, with $\lambda $ a constant, we
discover that the action (13) is just the action given in the expression (1)
of reference [27] (see also the expression (4.45) of reference [26]).
Therefore, we have shown that (8) is a different form to write (13). The
main goal in showing this equivalence is to be sure that (7) may provide a
possible starting point. At the same time, the procedure allowed us to
introduce some notation and useful relations which will become important in
the next sections.

\smallskip\ 

\smallskip\ 

\noindent {\bf 3.- \ 2-d GRAVITY AS ABELIAN GAUGE THEROY}

\smallskip\ 

We shall consider (8) assuming a vanishing torsion. In such a case, the
action (8) is reduced to

\begin{equation}
\begin{array}{c}
S=\frac{1}{4\alpha ^{2}}\int d^{2}x{}\sqrt{-g}(g^{\mu \alpha }g^{\nu \beta
}{}R_{\mu \nu }^{ab}{}R_{\alpha \beta }^{cd}{}\eta _{ac}\eta _{bd}+2g^{\mu
\alpha }g^{\nu \beta }{}\Sigma _{\mu \nu }^{ab}{}R_{\alpha \beta
}^{cd}{}\eta _{ac}\eta _{bd} \\ 
\\ 
+g^{\mu \alpha }g^{\nu \beta }{}\Sigma _{\mu \nu }^{ab}{}\Sigma _{\alpha
\beta }^{cd}{}\eta _{ac}\eta _{bd}).
\end{array}
\eqnum{14}
\end{equation}
At first sight this action looks as an action for a non-Abelian gauge field.
However, this is an illusion because in fact considering that $\omega _{\mu
}^{ab}=-\epsilon ^{ab}\omega _{\mu }$ the curvature (4) becomes

\begin{equation}
R_{\mu \nu }^{ab}=-\epsilon ^{ab}R_{\mu \nu },  \eqnum{15}
\end{equation}
where

\begin{equation}
R_{\mu \nu }=\partial _{\mu }\omega _{\nu }-\partial _{\nu }\omega _{\mu } 
\eqnum{16}
\end{equation}
and therefore the action (14) is reduced to

\begin{equation}
S=-\frac 1{\alpha ^2}\int d^2x{}\sqrt{-g}(\frac 12R^{\mu \nu }{}R_{\mu \nu
}+\epsilon ^{\mu \nu }{}R_{\mu \nu }-1),  \eqnum{17}
\end{equation}
which is an action associated to an Abelian gauge theory, with $\omega _\mu $
as a gauge field. The second term is the Euler number and classically \ do
not contribute to the dynamics. So, in a sense the action (17) is a
Maxwell-type action with field strength (16) and gauge field $\omega _\mu .$
The difference, however, is that, due to the vanishing of the torsion $%
T^a=de^a-\epsilon ^{ab}\omega \wedge e_b$, $\omega _\mu $ is related to the 
{\it Zweibein} field $e_\mu ^a$ by the formula $\omega _\mu =-\epsilon
^{\alpha \beta }\partial _\alpha e_\beta ^ae_{\mu a}$ (see Ref. [28])$.$

At this stage, it is convenient to rescale the {\it Zweibein} field $e_\mu
^a $ in the form $\lambda e_\mu ^a$, where $\lambda $ is a constant$.$ So
the metric changes to $g_{\mu \nu }\rightarrow \lambda ^2g_{\mu \nu }$ and
therefore $\sqrt{-g}\rightarrow \lambda ^2\sqrt{-g}$ and $g^{\mu \nu
}\rightarrow \lambda ^{-2}g^{\mu \nu }.$ We also need to rescale the
curvature as $R_{\mu \nu }\rightarrow \frac 1GR_{\mu \nu }$ where $G$ is
also a constant. Considering these rescalings the action (17) can be written
as

\begin{equation}
S=-\int d^2x{}\sqrt{-g}(\text{ }\frac 1{2G^2\lambda ^2\alpha ^2}R^{\mu \nu
}{}R_{\mu \nu }+\frac 1{G\alpha ^2}\epsilon ^{\mu \nu }{}R_{\mu \nu }-\frac{%
\lambda ^2}{\alpha ^2}).  \eqnum{18}
\end{equation}
Of course, $G_N=G\alpha ^2$ and $\Lambda =\frac{\lambda ^2}{\alpha ^2}$ can
be identified with the gravitational Newton constant and cosmological
constant in two dimensions, respectively. For our purpose, it turns out more
convenient to use the constants $\gamma =G\lambda \alpha $ and $\theta =%
\frac 1{G\alpha ^2}$ which, in analogy to the four dimensional Maxwell
theory, can be identified as the gravitational ''electric'' coupling
constant and the $\theta $ parameter respectively. Thus, we have

\begin{equation}
S=-\int d^2x{}\sqrt{-g}(\frac 1{2\gamma ^2}R^{\mu \nu }{}R_{\mu \nu }+\theta
\epsilon ^{\mu \nu }{}R_{\mu \nu }-\Lambda ).  \eqnum{19}
\end{equation}
It turns out that this action is equivalent to the action

\begin{equation}
I=-\frac 14\int d^2x{}\sqrt{-g}(^{+}\tau ^{+}R^{\mu \nu }{}^{+}R_{\mu \nu
}+^{-}\tau ^{-}R^{\mu \nu }{}^{-}R_{\mu \nu }),  \eqnum{20}
\end{equation}
where

\begin{equation}
^{\pm }R_{\mu \nu }=R_{\mu \nu }\pm \tilde{\lambda}\epsilon _{\mu \nu } 
\eqnum{21}
\end{equation}
and

\begin{equation}
^{\pm }\tau =\frac 1{\gamma ^2}\pm \frac \theta {\tilde{\lambda}}, 
\eqnum{22a}
\end{equation}
with $\tilde{\lambda}=\frac{\lambda \gamma }\alpha .$ If we give to $^{\pm
}\tau $ the typical complex form of a modular transformation

\begin{equation}
^{\pm }\tau =\frac 1{\gamma ^2}\pm \frac{i\theta }{\tilde{\lambda}}, 
\eqnum{22b}
\end{equation}
we note that (20) also leads to (19) but with the second term now as a
complex quantity $i\theta \epsilon ^{\mu \nu }{}R_{\mu \nu }.$

\smallskip\ 

\noindent {\bf 4.- S-DUALITY FOR 2-d GRAVITY}

\smallskip\ 

The computation of the S-dual 2-d gravitational action is now
straightforward. Let us first introduce the two form $G$ and let us set 
\begin{equation}
^{\pm }P_{\mu \nu }\equiv \text{ }^{\pm }R_{\mu \nu }-\text{ }^{\pm }G_{\mu
\nu },  \eqnum{23}
\end{equation}
with $^{\pm }G_{\mu \nu }=-$ $^{\pm }G_{\nu \mu }$.

Consider the partition function

\begin{equation}
Z=\int d^{+}G{}d^{-}G{}d\omega {}de{}dV{}e^{-I_E},  \eqnum{24}
\end{equation}
where $I_E$ is the extended action 
\begin{equation}
\begin{array}{ccc}
I_E= & -\frac 14\int d^2x\sqrt{-g}(^{+}\tau ^{+}P^{\mu \nu }{}^{+}P_{\mu \nu
}+^{-}\tau ^{-}P^{\mu \nu }{}^{-}P_{\mu \nu }) &  \\ 
&  &  \\ 
& -\frac 12\int d^2x\sqrt{-g}(^{+}W^{\mu \nu }{}^{+}G_{\mu \nu }+^{-}W^{\mu
\nu }{}^{-}G_{\mu \nu }). & 
\end{array}
\eqnum{25}
\end{equation}
Here, $^{\pm }W_{\mu \nu }=\partial _\mu V_\nu -\partial _\nu V_\mu $ $\pm 
\tilde{\lambda}\epsilon _{\mu \nu }$ is the dual field strength. The tensor $%
\Xi _{\mu \nu }=\partial _\mu V_\nu -\partial _\nu V_\mu $ satisfies the
Dirac quatization law 
\begin{equation}
\int \Xi \in 2\pi {\bf Z.}  \eqnum{26}
\end{equation}
We note that the partition function $Z$ is invariant not only under the
typical gauge invariance

\begin{equation}
\omega \rightarrow \omega +d\lambda ,G\rightarrow G  \eqnum{27}
\end{equation}
transformation, but also under 
\begin{equation}
\omega \rightarrow \omega +B\;and\;G\rightarrow G+dB,  \eqnum{28}
\end{equation}
where $B$ is an arbitrary one form.

We first need to show that (25) is equivalent to (20). For this purpose, let
us consider the partition function $Z$ containing the dual field $V$: 
\begin{equation}
\int DV\exp (-\frac{1}{2}\int d^{2}x\sqrt{-g}(^{+}W^{\mu \nu }{}^{+}G_{\mu
\nu }+^{-}W^{\mu \nu }{}^{-}G_{\mu \nu }).  \eqnum{29}
\end{equation}
Integrating out the dual connection $V$ at the classical level, we get a
delta function setting $dG=0.$ Thus, the gauge invariance (28) allows us to
set $G=0,$ reducing (25) to the original action (20). Therefore, the actions
(25) and (20) are, in fact, classically equivalents.

We next note that the gauge invariance (28) enables one to fix a gauge with $%
\omega =0.$ The action (25) is then reduced to

\begin{equation}
\begin{array}{ccc}
I_E= & -\frac 14\int d^2x\sqrt{-g}(^{+}\tau ^{+}Q^{\mu \nu }{}^{+}Q_{\mu \nu
}+^{-}\tau ^{-}Q^{\mu \nu }{}^{-}Q_{\mu \nu }) &  \\ 
&  &  \\ 
& -\frac 12\int d^2x\sqrt{-g}(^{+}W^{\mu \nu }{}^{+}G_{\mu \nu }+^{-}W^{\mu
\nu }{}^{-}G_{\mu \nu }), & 
\end{array}
\eqnum{30}
\end{equation}
where 
\begin{equation}
^{\pm }Q_{\mu \nu }\equiv \text{ }^{\pm }\Omega _{\mu \nu }-\text{ }^{\pm
}G_{\mu \nu },  \eqnum{31}
\end{equation}
with 
\begin{equation}
^{\pm }\Omega _{\mu \nu }=\pm \tilde{\lambda}\epsilon _{\mu \nu }. 
\eqnum{32}
\end{equation}
By virtue of (31) the extended action (30) becomes

\begin{equation}
\begin{array}{c}
I_E=-\frac 14\int d^2x\sqrt{-g}(^{+}\tau ^{+}G^{\mu \nu }{}^{+}G_{\mu \nu
}+^{-}\tau ^{-}G^{\mu \nu }{}^{-}G_{\mu \nu }^{+}) \\ 
\\ 
-\frac 14\int d^2x\sqrt{-g}(^{+}\tau ^{+}\Omega ^{\mu \nu }{}^{+}\Omega
_{\mu \nu }+^{-}\tau ^{-}\Omega ^{\mu \nu }{}^{-}\Omega _{\mu \nu }) \\ 
\\ 
-\frac 12\int d^2x\sqrt{-g}(^{+}L^{\mu \nu }{}^{+}G_{\mu \nu }+^{-}L^{\mu
\nu }{}^{-}G_{\mu \nu }),
\end{array}
\eqnum{33}
\end{equation}
where

\begin{equation}
^{\pm }L^{\mu \nu }=\text{ }^{\pm }W^{\mu \nu }-\text{ }^{\pm }\tau \text{ }%
^{\pm }\Omega ^{\mu \nu }.  \eqnum{34}
\end{equation}

Using the partition function (24) with (33) as an action and integrating
over $^{+}G_{\mu \nu }$ and $^{-}G_{\mu \nu }$ we get

\begin{equation}
\begin{array}{c}
S_E=-\frac 14\int d^2x{}\sqrt{-g}((-\frac 1{^{+}\tau })^{+}L^{\mu \nu
}{}^{+}L_{\mu \nu }+{}(-\frac 1{^{-}\tau })^{-}L^{\mu \nu }{}^{-}L_{\mu \nu
}) \\ 
\\ 
-\frac 14\int d^2x\sqrt{-g}(^{+}\tau ^{+}\Omega ^{\mu \nu }{}^{+}\Omega
_{\mu \nu }+^{-}\tau ^{-}\Omega ^{\mu \nu }{}^{-}\Omega _{\mu \nu }).
\end{array}
\eqnum{35}
\end{equation}
Substituting (34) into (35) and using the fact that $^{\pm }W_{\mu \nu
}=\partial _\mu V_\nu -\partial _\nu V_\mu $ $\pm \tilde{\lambda}\epsilon
_{\mu \nu }=\partial _\mu V_\nu -\partial _\nu V_\mu +^{\pm }\Omega ^{\mu
\nu }$ we finally get the dual action

\begin{equation}
S_{D}=-\frac{1}{4}\int d^{2}x{}\sqrt{-g}((-\frac{1}{^{+}\tau })^{+}W^{\mu
\nu }{}^{+}W_{\mu \nu }+(-\frac{1}{^{-}\tau })^{-}W^{\mu \nu }{}^{-}W_{\mu
\nu })+2\tilde{\lambda}^{2}\int d^{2}x{}\sqrt{-g}.  \eqnum{36}
\end{equation}
Note that the action (36) has the dual gauge invariance $V\rightarrow
V+d\alpha $ and, up to cosmological constant term, is of the general type
(20) but with $\tau $ replaced by $-\frac{1}{\tau },$ where $\tau $ can be
either $^{+}\tau $ or $^{-}\tau .$ Therefore, we have shown that the
coupling transforms as $\tau \rightarrow -\frac{1}{\tau }$ when one changes
from the action (20) to the action (36)$.$

\smallskip\ 

\noindent {\bf 5.-FINAL COMMENTS}

\smallskip\ 

In this article, we have shown that it is possible to associate the analogue
of S-duality to 2-d gravity. The main observation is that 2-d gravity can
have an interpretation of an Abelian gauge theory and therefore one can
apply standard procedure to find the S-dual action. Specifically, we proved
that the partition function for 2-d gravity has the exact S-dual symmetry

\begin{equation}
Z_\omega (\tau )=Z_V(-\frac 1\tau ),  \eqnum{37}
\end{equation}
where $Z_\omega (\tau )$ is the partition function associated to the action
(20), while $Z_V(-\frac 1\tau )$ is the partition function associated to the
action (36).

A natural question is now to investigate what is the strong coupling limit
of 2-d gravity. In this limit we should expect a decompactified dimension.
So, we are searching for a gravitational theory in 2+1 dimensions. But as it
was mentioned in the introduction from Chern-Simons theory it is possible to
obtain the action of Jackiw-Teitelboim and
Callan-Giddings-Harvey-Strominger, which are not quadratic in the curvature.
On the other hand, it is known that 2+1 dimensional gravity can be obtained
from Euler and Pontrjagin topological invariant terms in 3+1 dimensions.
Therefore, since this two topological invariants can be written as a
Chern-Simons actions in 2+1 dimensions one should expect that we must go to
3+1 dimensions but with an additional term quadratic in the curvature.
Fortunately, topological gravity in 3+1 dimensions seems to accomplish the
required combinations in the curvature. In fact, a typical form of 3+1
topological gravity is a term of the form

\begin{equation}
S=-\int d^4x{}\sqrt{-g}\frac 1{2\gamma ^2}(R^{\hat{\mu}\hat{\nu}}+\frac 12%
\epsilon ^{\hat{\mu}\hat{\nu}\hat{\alpha}\hat{\beta}}{}R_{\hat{\alpha}\hat{%
\beta}})(R_{\hat{\mu}\hat{\nu}}+\frac 12\epsilon _{\hat{\mu}\hat{\nu}%
}^{\quad \hat{\alpha}\hat{\beta}}{}R_{\hat{\alpha}\hat{\beta}})  \eqnum{38}
\end{equation}
which under dimensional reduction one should expect to lead to the 2-d
gravitational action (8). Thus, this roughly observations lead us to
conjecture that the strong coupling limit of 2-d gravity must be (something
like) topological gravity in 3+1 dimensions. It is interesting that our
conclusion is in agreement with stochastic quantization theory where a
similar connection between 2-d gravity and 4-d topological gravity is found
[29].

So far, our discussion has focused on pure 2-d gravity. But what about 2-d
gravity interacting with matter fields. For the case of a scalar field we do
not even need to consider an action with quadratic curvature to find the
analogue of S-dual for 2-d gravity. In fact, consider the action [30]

\begin{equation}
S=-\int d^2x{}\sqrt{-g}(\kappa D^\mu \varphi D_\mu \varphi +\Lambda ^{\mu
\nu }{}R_{\mu \nu }),  \eqnum{39}
\end{equation}
where

\begin{equation}
D_\mu =\partial _\mu +\omega _\mu ,  \eqnum{40}
\end{equation}
$\kappa $ is a constant and $\Lambda ^{\mu \nu }$ is a Lagrange multiplier.
If we perform an integration with respect the Lagrange multiplier $\Lambda
^{\mu \nu }$ we obtain $R_{\mu \nu }=0.$ Therefore, this result together
with the gauge invariance $\omega _\mu \rightarrow \omega _\mu +\partial
_\mu f$ allows us to set $\omega _\mu =0,$ reducing (39) to

\begin{equation}
S=-\kappa \int d^2x{}\sqrt{-g}\partial ^\mu \varphi \partial _\mu \varphi 
\eqnum{41}
\end{equation}
which is the typical action for a scalar field .

Let us go back to (39). Since (39) is also invariant under $\varphi
\rightarrow \varphi +b$ and $\omega \rightarrow \omega +db$ we can set $%
\varphi =0.$ Integrating by parts the resultant action and integrating out $%
\omega $ we get the dual action for the variable $\Phi =\frac{1}{2}%
\varepsilon _{\mu \nu }\Lambda ^{\mu \nu }$;

\begin{equation}
S=-(-\frac 1\kappa )\int d^2x{}\sqrt{-g}\partial ^\mu \Phi \partial _\mu
\Phi ,  \eqnum{42}
\end{equation}
showing the typical dual transformation $\kappa \rightarrow -\frac 1\kappa $
for the coupling constant$.$

For the case of 2-d supergravity [31] and [28] we can write the partition
function for 2-d gravity as a product of two partition functions: one
corresponding to 2-d gravity and the other corresponding to Rarita-Schwinger
field. This way, one should expect to get the S-dual 2-d supergravity
symmetry $Z_{\omega ,\psi }(\tau )=Z_{V,\varphi }(-\frac{1}{\tau }),$ where $%
\psi $ is the Rarita-Shwinger field and $\varphi $ its dual.

Some time ago, Ikeda and Izawa [32] (see also [33]) used nonlinear
Poincar\'{e} algebra to construct a gauge theory for 2-d gravity which turns
out to be equivalent to most general Poincar\'{e} gauge theory for 2-d
gravity with dynamical torsion:

\begin{equation}
S=\int d^2x{}\sqrt{-g}(\frac 1{4\alpha ^2}g^{\mu \alpha }g^{\nu \beta
}{}R_{\mu \nu }^{ab}{}R_{\alpha \beta }^{cd}{}\eta _{ac}\eta _{bd}+\frac 1{%
2\beta ^2}g^{\mu \alpha }g^{\nu \beta }{}T_{\mu \nu }^a{}T_{\alpha \beta
}^b{}\eta _{ab}+\Lambda ).  \eqnum{43}
\end{equation}
Here, $\alpha ,\beta ,$and $\Lambda $ are different constants. Clearly,
dropping from this action the torsion and including the $\theta $ term we
obtain our action (19). In a similar way, it has been shown that the
quadratic W$_3$ algebra leads to the W$_3$ gravity [34]. So it seems that
from our work follows that it must also be possible to associate the
analogue of S-duality to W$_3$ gravity.

Another interesting observation is that when a 4-d Abelian gauge theory is
coupled to gravity, it has been found that the partition function is not a
modular-invariant function but transform as a modular form [7]. In view of
the result of the present work it seems that in two dimensions the partition
function of an Abelian gauge field coupled to gravity must be
modular-invariant function.

Finally, it is known that Liouville 2-d gravity theory has a very
interesting features in the strong coupling regime [35]. It may be
interesting to understand such a features from the point of view of the
present work. It is also known that the S-duality gauge invariance is deeply
related to bosonization [16], 2-d ADS / CFT correspondence [36], and
noncommutative gauge theory [37]. It may be also very interesting to see
whether the present work can be useful in those directions.

\smallskip\

\end{document}